\documentclass[]{spie}  %>>> use for US letter paper
\usepackage{amsmath,amsfonts,amssymb}
\usepackage{graphicx}
\usepackage[colorlinks=true, allcolors=blue]{hyperref}
\graphicspath{ {images/} }
\usepackage[skip=10pt]{caption} % example skip set to 2pt
\usepackage{subcaption}
\usepackage[export]{adjustbox}
\usepackage{pdfpages}
\usepackage{placeins}
\usepackage{xcolor}
\usepackage[utf8]{inputenc}
\usepackage{algpseudocode}
\usepackage{algorithm}
\usepackage{tabularx}
\usepackage{pdflscape}

\usepackage{float}

\title{From SuperBIT to GigaBIT: Informing next-generation balloon-borne telescope design with Fine Guidance System flight data}

\author[a]{Philippe Voyer}
\author[b]{Steven J.\ Benton}
\author[a]{Christopher J.\ Damaren}
\author[c]{Spencer W.\ Everett}
\author[b]{Aurelien A.\ Fraisse}
\author[d]{Ajay S.\ Gill}
\author[e]{John W.\ Hartley}
\author[f]{David Harvey}
\author[a]{Michael Henderson}
\author[a]{Bradley Holder}
\author[c]{Eric M.\ Huff}
\author[g,h]{Mathilde Jauzac}
\author[b]{William C.\ Jones}
\author[g,h]{David Lagattuta}
\author[i]{Jason S.-Y.\ Leung}
\author[e]{Lun Li}
\author[b]{Thuy Vy T.\ Luu}
\author[g,h]{Richard Massey}
\author[j]{Jacqueline E.\ McCleary}
\author[k]{Johanna M.\ Nagy}
\author[i,l]{C.\ Barth Netterfield}
\author[l]{Emaad Paracha}
\author[c,m]{Susan F.\ Redmond}
\author[c]{Jason D.\ Rhodes}
\author[c]{Andrew Robertson}
\author[e]{L.\ Javier Romualdez}
\author[g]{J\"urgen Schmoll}
\author[n]{Mohamed M.\ Shaaban}
\author[o]{Ellen L.\ Sirks}
\author[j]{Georgios N.\ Vassilakis}
\author[c]{Andr\'{e} Z.\ Vitorelli}

\affil[a]{University of Toronto Institute for Aerospace Studies (UTIAS), 4925 Dufferin Street, Toronto, ON, Canada}

\affil[b]{Department of Physics, Princeton University, Jadwin Hall, Princeton, NJ, USA}

\affil[c]{Jet Propulsion Laboratory, California Institute of Technology, 4800 Oak Grove Drive, Pasadena, CA, USA}

\affil[d]{Department of Aeronautics and Astronautics, Massachusetts Institute of Technology, 77 Massachusetts Ave, Cambridge, MA 02139 USA}

\affil[e]{StarSpec Technologies Inc. Unit C-5, 1600 Industrial Road, Cambridge, ON, Canada, N3H 4W5}

\affil[f]{Laboratoire d’Astrophysique, EPFL, Observatoire de Sauverny, 1290 Versoix, Switzerland}

\affil[g]{Institute for Computational Cosmology, Department of Physics, Durham University, South Road, Durham DH1 3LE, UK}

\affil[h]{Centre for Extragalactic Astronomy, Department of Physics, Department of Physics, Durham University, Durham DH1 3LE, UK}

\affil[i]{Dunlap Institute for Astronomy and Astrophysics, University of Toronto, 50 St. George Street, Toronto, ON, Canada M5S 3H4}

\affil[j]{Department of Physics, Northeastern University, 360 Huntington Ave, Boston, MA USA}

\affil[k]{Department of Physics, Case Western Reserve University, Rockefeller Bldg.,  10900 Euclid Ave, Cleveland, OH 44106}

\affil[l]{Department of Physics, University of Toronto, 60 St. George Street, Toronto, ON, Canada M5R 2M8}

\affil[m]{California Institute of Technology, 1216 E California Blvd, Pasadena, CA 91125, USA}

\affil[n]{Palantir Technologies 1875 Lawrence St, Denver, CO 80202, USA}

\affil[o]{School of Physics, The University of Sydney and ARC Centre of Excellence for Dark Matter Particle Physics, NSW 2006, Australia}

\authorinfo{Further author information: (Send correspondence to P.V.)\\P.V.: E-mail: philippe.voyer@mail.utoronto.ca}

\authorinfo{Further author information: (Send correspondence to P.V.)\\ E-mail: philippe.voyer@mail.utoronto.ca}

\begin{document} 
\maketitle
\newpage 
\begin{abstract}
The Super-pressure Balloon-borne Imaging Telescope (SuperBIT) is a near-diffraction-limited 0.5 m telescope that launched via NASA’s super-pressure balloon technology on April 16, 2023. SuperBIT achieved precise pointing control through the use of three nested frames in conjunction with an optical Fine Guidance System (FGS), resulting in an average image stability of 0.055" over 300-second exposures. The SuperBIT FGS includes a tip-tilt fast-steering mirror that corrects for jitter on a pair of focal plane star cameras. In this paper, we leverage the empirical data from SuperBIT's successful 45-night stratospheric mission to inform the FGS design for the next-generation balloon-borne telescope. The Gigapixel Balloon-borne Imaging Telescope (GigaBIT) is designed to be a 1.35m wide-field, high resolution imaging telescope, with specifications to extend the scale and capabilities beyond those of its predecessor SuperBIT. A description and analysis of the SuperBIT FGS will be presented along with methodologies for extrapolating this data to enhance GigaBIT's FGS design and fine pointing control algorithm. We employ a systems engineering approach to outline and formalize the design constraints and specifications for GigaBIT’s FGS. GigaBIT, building on the SuperBIT legacy, is set to enhance high-resolution astronomical imaging, marking a significant advancement in the field of balloon-borne telescopes. 
\end{abstract}

\\

\section{INTRODUCTION}
\label{sec:intro}  % \label{} allows reference to this section

The Super-pressure Balloon-borne Imaging Telescope (SuperBIT) is a near-diffraction-limited 0.5 m telescope in use since 2015 that launched via NASA’s super-pressure balloon technology in April 2023. SuperBIT achieved precise pointing control through the use of three nested frames in conjunction with an optical Fine Guidance System (FGS), resulting in an image stability of 0.055" over 300-second exposures. The success of the SuperBIT observatory in its recent science flight has prompted the development of a next-generation balloon-borne telescope, GigaBIT. The Gigapixel Balloon-borne Imaging Telescope (GigaBIT) is designed to be a 1.35m wide-field, high resolution imaging telescope, with specifications to extend the scale and capabilities beyond those of its predecessor SuperBIT. To meet their stringent pointing requirements, SuperBIT and GigaBIT are equipped with a Fine Guidance System (FGS) that employs precision opto-mechanical components, such as a Fast Steering Mirror (FSM), to correct for jitter and achieve sub-arcsecond image stability. 

The following focuses on the comprehensive modeling, control, and systems engineering of GigaBIT's FGS, incorporating insights and lessons learned from SuperBIT. Using FGS data from SuperBIT's 2023 science flight, feedback sensors are compared and analysed, and assumptions about the operating environment are made to inform the FGS control design. A high-fidelity integrated model of the SuperBIT FGS is developed using line-of-sight (LOS) ray-tracing theory, incorporating the flight data into the simulation framework as a disturbance. Various control strategies, including Proportional (P), Proportional-Integral-Derivative (PID), and Linear Quadratic Regulator (LQR) controls, are explored and evaluated for their effectiveness in managing jitter using a model-in-the-loop (MIL) approach. Finally, conclusions of this analysis drive the requirements for the FGS of GigaBIT as part of the Phase A model-based systems engineering (MBSE) study.\\

\section{BACKGROUND}

\subsection{SuperBIT Instrument}
\begin{figure}[]
        \centering
        \includegraphics[width = 0.65\textwidth]{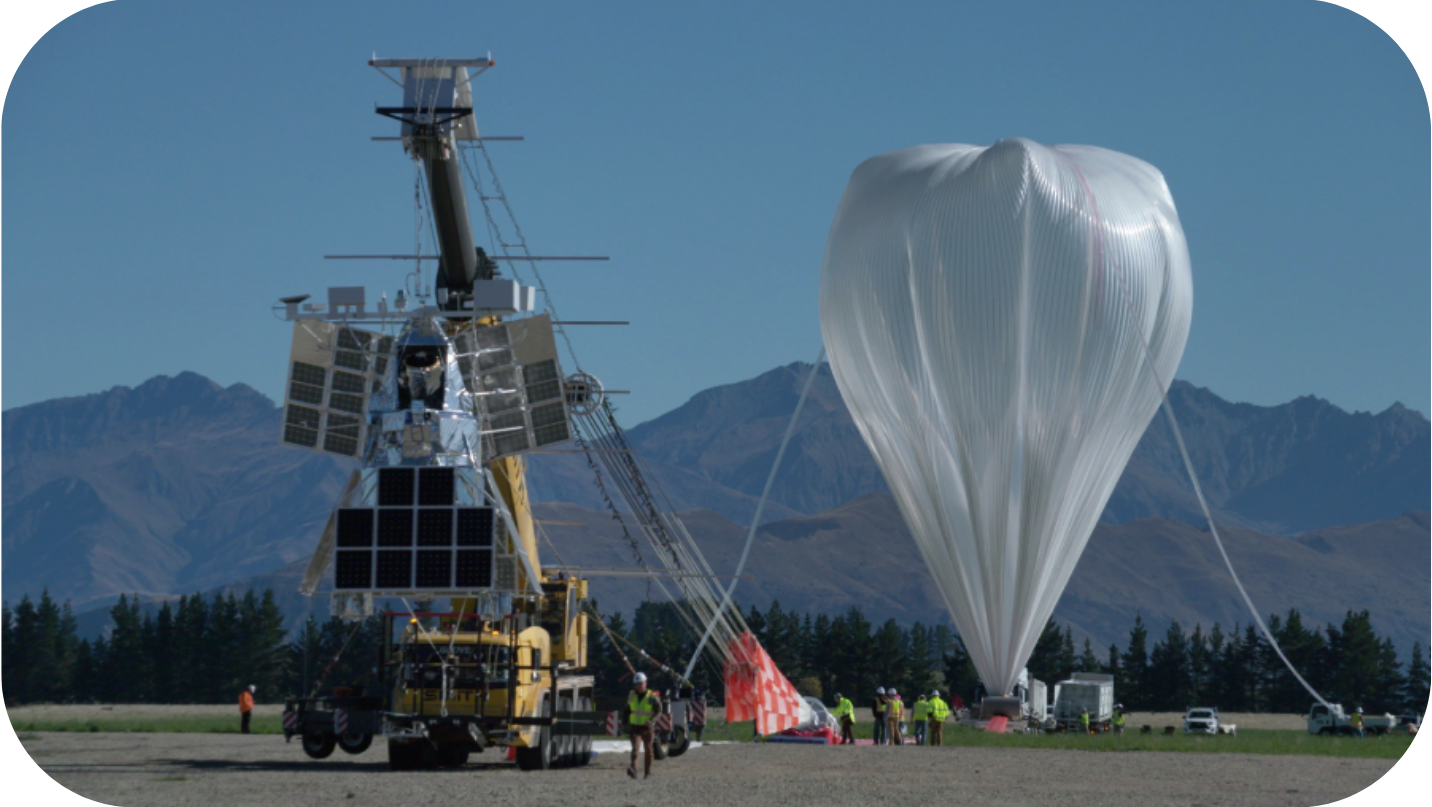}
        \caption{SuperBIT launch from Wānaka, New Zealand, in April 2023. Photo credit: NASA/Bill Rodman.} 
        \label{fig:SB_launchpad}
\end{figure}

The Super-pressure Balloon-borne Imaging Telescope (SuperBIT) was a 0.5-meter near-ultraviolet (NUV) to near-infrared (NIR) wide-field Dall-Kirkham style telescope that operated from 2015 to 2023.  Its initial phase included four test flights conducted between 2015 and 2019 \cite{Romualdez2019, Romualdez2020}, culminating in a science mission in 2023. The SuperBIT super-pressure science flight (flight 728NT), as seen in Figure \ref{fig:SB_launchpad}, launched from the W\=anaka airport in New Zealand on April 16, 2023 (11:42 PM UTC on April 15, 2023) on a NASA super-pressure balloon (SPB). The mission successfully concluded after 45 nights, consistently meeting its technical benchmarks, with preliminary analysis indicating diffraction-limited pointing. 
SuperBIT's 2023 science flight focused on high-resolution observations of 30 galaxy clusters to study the distribution of dark matter in galaxy clusters through weak gravitational lensing. 

From an engineering standpoint, SuperBIT is a three-axis stabilized telescope platform engineered to deliver sub-arcsecond pointing stability. A series of three nested frames (inner frame, middle frame and outer frame), actuate yaw, pitch and roll to meet the strict science-driven pointing requirements. The outer frame is controlled in yaw by the reaction wheel and pivot, the middle frame actuates roll about the outer frame, and the inner frame actuates telescope pitch about the middle frame. All three frames act as a three-axis stabilizer for the telescope, designed to  stabilize the telescope to within $1 - 2$". In addition, a back-end image stabilization subsystem further improves focal plane output to sub-acrsecond levels. \\

\subsubsection{SuperBIT Fine Guidance System}

\begin{figure}[H]
        \centering
        \includegraphics[width = 0.7\textwidth]{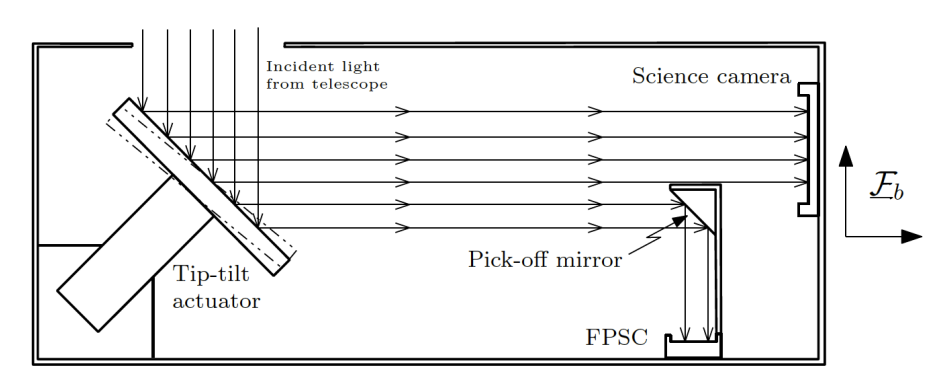}
        \caption{SuperBIT Fine Guidance System diagram \cite{Romualdez2018}.}
        \label{fig:SB_FGS}
\end{figure}

 The SuperBIT FGS, whose main purpose is jitter management, complements the three-axis gimbal system, further enhancing image stability to meet the strict pointing requirements. As schematized in Figure \ref{fig:SB_FGS}, the FGS is composed of a fast-steering mirror (FSM) assembly, two pick-off mirrors and two focal plane star cameras (FSCs) that provide sky-fixed feedback. The FSM, equipped with a flat 3-inch mirror mounted on a piezo-electric tip-tilt (TT) platform with a 2 milli-rad (mrad) throw (\textit{PI} S-340), addresses the jitter management control problem by minimizing star centroid movement on the FSCs.

\subsection{GigaBIT Instrument}
The Gigapixel Balloon-borne Imaging Telescope (GigaBIT) is designed to be a 1.35m wide-field, high resolution imaging telescope, with specifications to extend the scale and capabilities beyond those of its predecessor SuperBIT. It will provide 0.1 arc-second resolution imaging over a 0.18 degree squared field of view (FOV) with spectral coverage between 300 and 800 nm in 310 seconds of integration. The current telescope design for GigaBIT is a Three Mirror Anastigmatic (TMA) reflector supported by the gondola, which adopts the same three nested frames architecture from SuperBIT. GigaBIT is currently in its project Phase A, with the GigaBIT Request for Proposal (RFP) issued in 2024.\\

\subsubsection{FGS Control Problem}

\begin{figure}[H]
    \centering
    \includegraphics[width=0.8\textwidth]{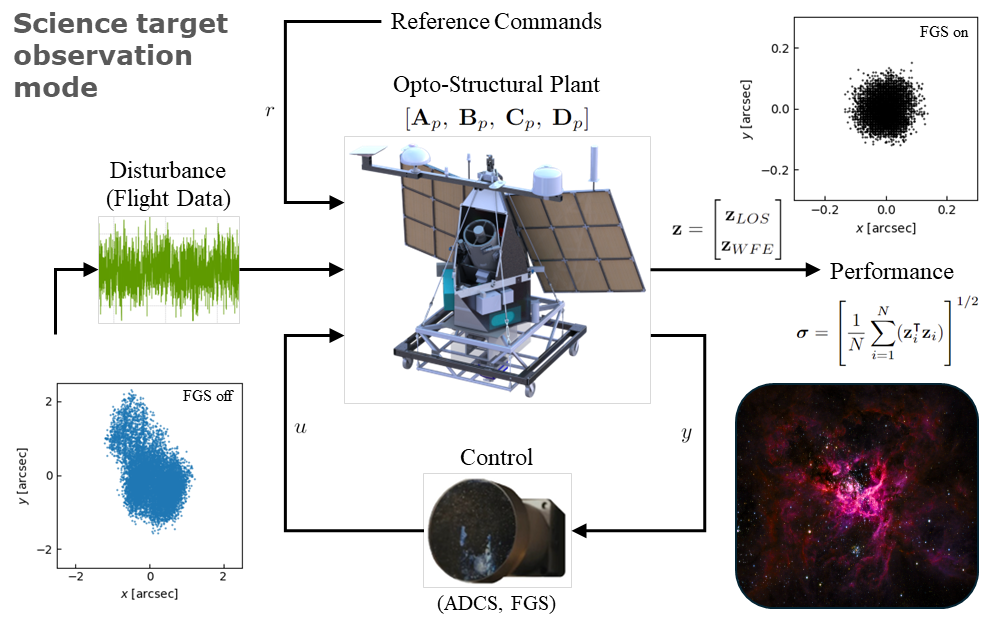}
    \caption[]{Reference problem as a generalized plant: Science target observation mode of an airborne telescope with pointing and phasing performances. Figure is inspired by Fig. 1.2 from de Weck et al.\cite{deWeck2000}}
    \label{fig:reference_problem}
\end{figure}

The substantial increase in size from SuperBIT to GigaBIT — nearly three times larger — poses a practical challenge in scaling the Fine Guidance System (FGS). This necessitates a thorough examination of the requirements and an initial analysis of the essential opto-mechanical systems for GigaBIT. The FGS control problem is summarized in Figure \ref{fig:reference_problem}, where the output LOS and wavefront (WFE) error are controlled given system disturbances.

To test the performance of the telescope prior to construction, and integration, the path from disturbance sources to the performance metrics of interest must be modeled. As such, the next section analyzes flight data to gain insight into the operating environment, which informs the development of a high-fidelity model of the SuperBIT FGS, with specifications scaled to the size of GigaBIT. \\

\section{FGS FLIGHT DATA ANALYSIS}

\begin{figure}[]
    \centering
    \includegraphics[width=\textwidth]{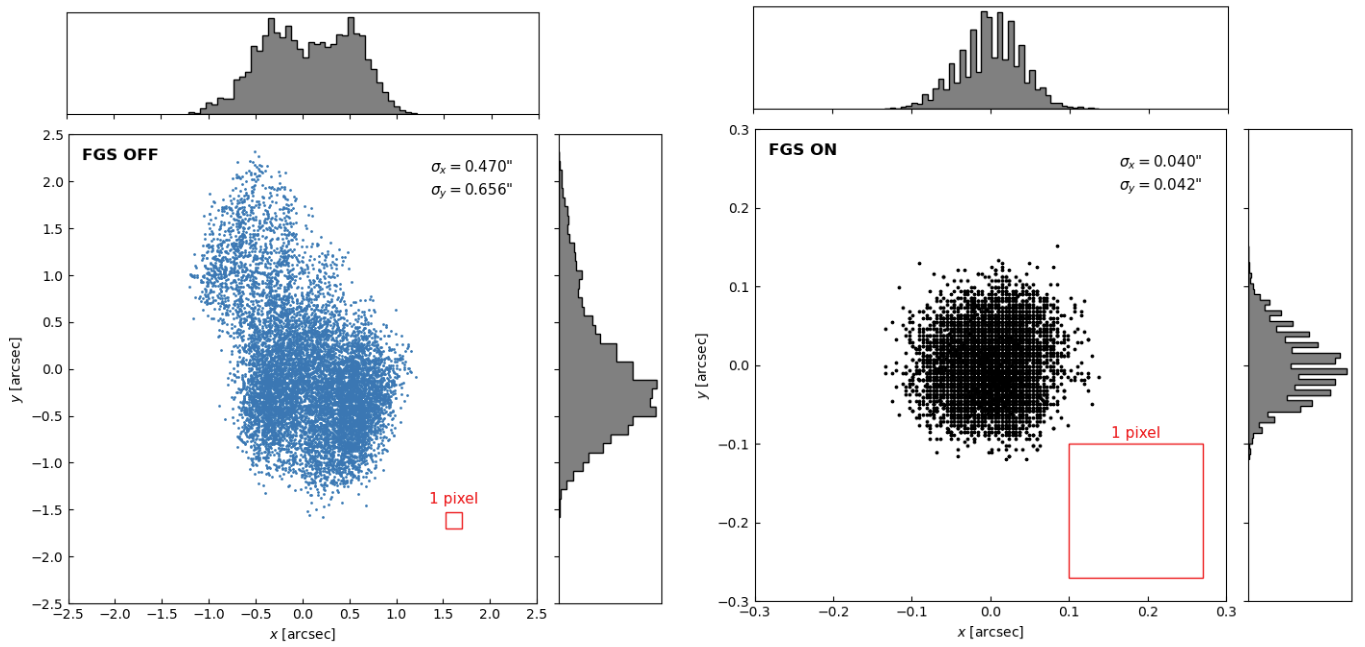}
    \caption[]{Raw 2023 FSC2 guide star centroid flight data with pixel scale $0.1703$ arcsec/px with FGS inactive for $\sim840$ seconds (left) and FGS active for $\sim466$ seconds (right).}
    \label{fig:scatterplots}
\end{figure}

From Figure \ref{fig:scatterplots}, the FGS performance in managing jitter is made evident, with a marked improvement in both $x$ and $y$ centroid standard deviation of almost $10\times$. The left FGS-off dataset displays a scatter of $10080$ centroiding outputs for $\sim840$ seconds of jitter data, implying an average FSC2 sampling rate of $12$ Hz over this time interval. In contrast, the FGS-on dataset implies an average centroid output rate of $14.1$ Hz for the time interval of $460$ seconds. This suggests that the FSC, with its $10-20$ Hz sampling rate, introduces latency in providing measurements to the controller, thus it should be complemented by other sensors for feedback. Since the centroiding algorithm is influenced by the brightness of the guide star, the camera may take longer to process measurements than other sensors. Also note that the grid effect seen in the right plot of Figure \ref{fig:scatterplots} is due to the centroiding algorithm that assigns the $x$ and $y$ positions to the closest $1/10$ of a pixel, which is sufficient for imaging purposes.

By analyzing the star centroid data with the FGS inactive, several observations can be made. Firstly, the focal plane star camera (FSC1 and FSC2) solutions can be compared against the integrated iXBlue gyroscopes. Secondly, this analysis provides insights into the operating environment and the level of jitter that must be mitigated by the control system.

\begin{figure}[]
    \centering
    \includegraphics[width=\textwidth]{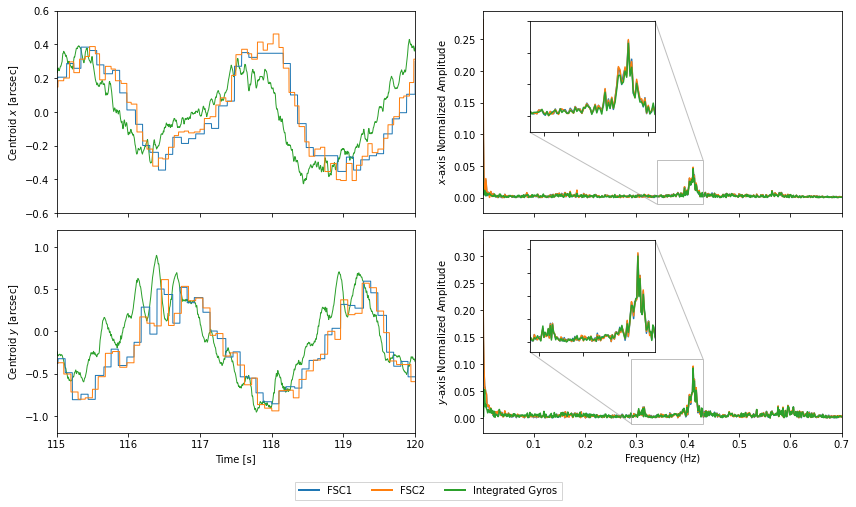}
    \caption[]{Comparison centroid position over time (left) and FFT amplitude spectrum (right) for different sensors from SuperBIT 2023 flight.}
    \label{fig:FGS_off_data}
\end{figure}

In terms of sensor performance, the FSCs, which provide the most accurate position of the star centroid, inherently exhibit data latency as a result of the required longer exposure times necessary for accurate star detection. This latency manifests as a zero-order hold as seen in the left plot of Figure \ref{fig:FGS_off_data}, leading to the observed step-like structure in the FSC traces. Both FSCs demonstrate closely matching curves, indicating a high degree of correlation in their measurements. The iXBlue gyroscopes, despite not directly measuring the star centroid position, offer high-resolution data reflective of spacecraft jitter. The gyros, with a high sampling rate, provide a continuous assessment of motion, which can be correlated with the centroid position to infer the star's apparent movement. This is evidenced by the smoother trend lines in the gyro data within the figures. Additionally, the left plot of Figure \ref{fig:FGS_off_data} presents a phase shift between the data from the iXBlue gyroscopes and the FSCs, indicating a time lag that can be attributed to the inherently slower response of the FSCs. The gyroscopes, in contrast, update at 300 Hz, capturing the spacecraft's motion without such delays. This difference in response times is an important consideration for integrating the data from these sensors.

The Fast Fourier Transform (FFT) analysis of the telescope's focal plane jitter, shown on the right plot of Figure \ref{fig:FGS_off_data}, reveals a prominent peak at $0.4108$ Hz, attributed to the pendulation of the balloon environment, with the peak's larger amplitude in the $y$-axis data due to alignment with the yaw direction where pendulation is most significant. This pendulatory motion is consistently captured by all three sensors -- FSC1, FSC2, and the integrated gyros -- indicating a common influence rather than isolated disturbances. Notably, FSC1 and FSC2 exhibit spectral leakage near $0$ Hz, likely from data drift, which can be ignored. Additional smaller peaks at around $0.31$ Hz and $0.6$ Hz in the $y$-axis are also observed. Importantly, all critical frequencies needing stabilization fall below $1$ Hz, favorably impacting the design of the control system by confining the required response range to lower frequencies where actuation and correction can be more efficiently managed.

\subsection{State Estimation Discussion}
The primary problem here is to effectively manage the inherent latency in the FSCs. This boils down to a state estimation problem, where a mix of rate gyroscope and FSC are used for feedback. With the assumption of linearity between the FSM axes, a state estimation algorithm was developed for the SuperBIT FGS using the FSC readout with rate gyroscope compensation to determine the centroid state and covariance at every time step. This procedure is detailed in Chapter 5.5 of Romualdez's PhD thesis\cite{Romualdez2018}. Romualdez outlines a Kalman Filter that adjusts the position estimates from the FSC by accounting for the time step between the gyroscope measurements and the FSC's data, essentially predicting the FSC's updated position. Finally, this is further refined with a weighting term that reflects the confidence in the gyroscopes’ data. The best option for GigaBIT is to employ a similar sensor fusion approach, integrating both FSC and gyroscope data to achieve optimal state estimation and feedback performance.

\section{FGS MODELING}

\subsection{LOS Kinematic Model}
To model the back-end image stabilization system in simulation, a LOS ray-tracing model is first defined. It uses fundamental ray-tracing theory first outlined by Silbertein \cite{Silberstein1918} and later derived in more detail by Redding and Breckenridge \cite{Redding1991}.

In short, and for practical implementation, a {\small \texttt{REFLECTION}} function was written to streamline the LOS model. This algorithm inputs the control ($\theta$, $\phi$) and mounting angles ($\alpha$, $\beta$), two points specific to the optical layout, $\textbf{p}_0$ and $\textbf{p}_r$, as well as the direction of the incident ray, $\textbf{r}_i$. It then processes the equations outlined in Algorithm 1, yielding both the point where the ray intersects the optical surface, $\textbf{p}_{si}$ and the direction of the reflected ray, $\textbf{r}_r$, as outputs. As such, this function should be called one time for every reflective surface in the optical layout of interest. 

\begin{figure}[]
    \centering
    \captionof{table}{Pseudocode for the ray-tracing intersect function.}
    \includegraphics[width=0.87\textwidth]{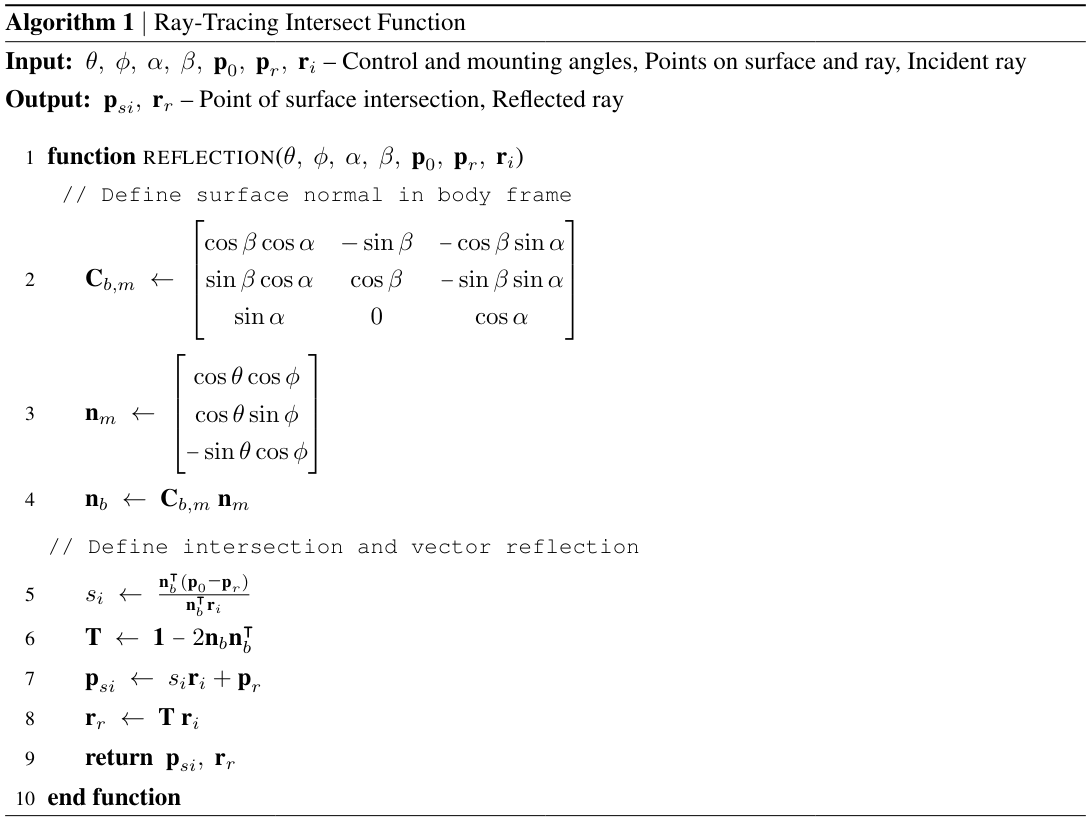}
    \label{fig:reflection}
\end{figure}

\begin{figure}[]
    \centering
    \includegraphics[width=0.75\textwidth]{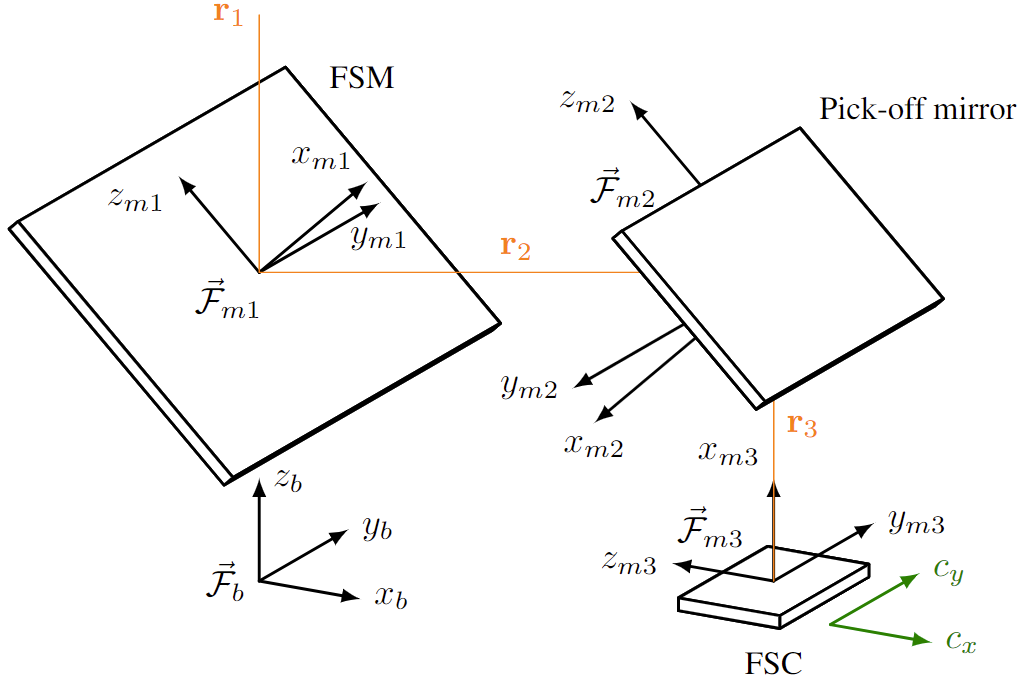}
    \caption[]{SuperBIT FGS Model: Illustrates the surfaces and reference frames of the FSM, pick-off mirror, and FSC, featuring ray-tracing LOS in orange and image frame in green.}
    \label{fig:FGS_frames}
\end{figure}

To simulate the LOS ray-tracing model, the SuperBIT FGS opto-mechanical design is used as a case study. Figure \ref{fig:FGS_frames} shows the optical surface reference frames used for the model as a near-orthogonal view of the Figure \ref{fig:SB_FGS} diagram. As such, the {\small \texttt{REFLECTION}} function was called three times with the SuperBIT FGS specifications as inputs. The result is the nonlinear kinematic relationship between the $x$ and $y$ position of the centroid on the image frame and the FSM tip and tilt angles, $\theta$ and $\phi$:

\begin{equation}
    \textbf{c} = \begin{bmatrix}
        c_x (\theta,\phi) \\ c_y(\theta,\phi)
    \end{bmatrix}.
    \label{eq:centroid}
\end{equation}

\noindent where $c_x$ and $c_y$ are rather long and nonlinear functions of tip and tilt angles. To simplify, the equations were then linearized about the $(0,0)$ equilibrium point to yield:

\begin{equation}
     \textbf{c}_{L} = \begin{bmatrix}
        0.337915\theta \\ 0.220074\phi
    \end{bmatrix}= \begin{bmatrix}
        0.337915 & 0 \\ 0 & 0.220074
    \end{bmatrix}\begin{bmatrix}
        \theta \\ \phi
    \end{bmatrix}=\underbrace{\begin{bmatrix}
        J_x & 0 \\ 0 & J_y
    \end{bmatrix}}_{\textbf{J}}\begin{bmatrix}
        \theta \\ \phi
    \end{bmatrix}
    \label{eq:lin_model}
\end{equation}

\noindent where units are [m] for positions and [rad] for angles. To generalize, $J_x$ and $J_y$ serve as conversion factors, allowing for flexibility on desired units. $\textbf{J}$ can also be thought of as a sensitivity matrix that describes the rate of change of the system's outputs with respect to its inputs. 

\subsection{Dynamic State-Space Model}

Next, a dynamic state-space model is defined, inspired by the work of Watkins and Agrawal \cite{Watkins2007}. The state-space model, with $\textbf{x}=\begin{bmatrix} \theta\ \phi\ \dot{\theta}\ 
 \dot{\phi} \end{bmatrix}^\intercal$ and $\textbf{y} = \begin{bmatrix} c_x\  c_y \end{bmatrix}^\intercal$, is given as follows:

\begin{equation}
\begin{bmatrix} \dot{\theta} \\ \ddot{\theta} \\ \dot{\phi} \\ \ddot{\phi} \end{bmatrix} =
\underbrace{\begin{bmatrix}
0 & 1 & 0 & 0 \\
-\omega_{x}^2 & -2\zeta_x \omega_{x} & A_x\omega_{x}^2 & A_x 2\zeta_x \omega_{x} \\
0 & 0 & 0 & 1 \\
A_y\omega_{y}^2 & A_y 2\zeta_y \omega_{y} & -\omega_{y}^2 & -2\zeta_y \omega_{y} \\
\end{bmatrix}}_\textbf{A}
\begin{bmatrix} \theta \\ \dot{\theta} \\ \phi \\ \dot{\phi} \end{bmatrix}
+
\underbrace{\begin{bmatrix}
0 & 0 \\ 
 \omega_{x}^2 & 0 \\
0 & 0 \\ 
0 & \omega_{y}^2 \\ 
\end{bmatrix}}_\textbf{B} \begin{bmatrix}
    u_1 \\ u_2
\end{bmatrix},
\end{equation}

\begin{equation}
    \textbf{y} = \textbf{J}\ \begin{bmatrix} 1 & 0 & 0 & 0 \\ 0 & 0 & 1 & 0 \end{bmatrix}= \begin{bmatrix}
        J_x & 0 \\ 0 & J_y
    \end{bmatrix} \begin{bmatrix}
        \theta \\ \phi
        \end{bmatrix} = \underbrace{\begin{bmatrix} J_x & 0 & 0 & 0  \\ 0 & 0 & J_y & 0\end{bmatrix}}_\textbf{C} \begin{bmatrix} \theta \\ \dot{\theta} \\ \phi \\ \dot{\phi} \end{bmatrix}
\end{equation}

\noindent where $J_x$ and $J_y$ serve as conversion factors to accommodate for unit changes from the selected inputs, as described by the previously defined linear Eq. (\ref{eq:lin_model}). The other parameters, some taken from the SuperBIT hardware specifications, are listed in Table \ref{tab:SSparameters}. To scale the model from SuperBIT to GigaBIT, the natural frequency was multiplied by a factor of $1.5$, resulting in a slightly slower response due to the expected larger FSM size.

\begin{table}[H]
\caption{State-space model parameter values.}
\centering
\begin{tabular}{lcc}
\hline 
Name                                & Variable                       & Value                         \\ \hline
Tip rotation about $x$ axis of FSM  & $\theta$                       & $-2$ to $+2$ mrad                 \\
Tilt rotation about $y$ axis of FSM & $\phi$                         & $-2$ to $+2$ mrad                 \\
FSM damping ratio, $x$ axis         & $\zeta_x$                     & 0.7                           \\
FSM damping ratio, $y$ axis         & $\zeta_y$                     & 0.7                           \\
FSM natural frequency, $x$ axis     & $\omega_x$                    & $700$ rad/s        \\
FSM natural frequency, $y$ axis     & $\omega_y$                    & $700$ rad/s            \\
Cross-coupling factor, $x$ axis     & $A_x$                          & $-2\times10^{-2}$ \\
Cross-coupling factor, $y$ axis     & $A_y$                          & $-2\times10^{-2}$ \\
Conversion factor, $x$ axis & $J_x$ & $13141$ "/rad ($0.337915$ m/rad) \\
Conversion factor, $y$ axis & $J_y$ & $8558$ "/rad ($0.220074$ m/rad) \\ \hline 
\end{tabular}
\label{tab:SSparameters}
\end{table}

This dynamic model's response is concurrent with the \textit{PI} S-340 manufacturer report, with a step input settling time of $5$ ms.\\

\section{CONTROL DESIGN}
Using the previously defined model of the FGS, different control strategies are tested to compare their performance at managing jitter. Specifically, Proportional (P), Proportional-Integral-Derivative (PID), and Linear Quadratic Regulator (LQR) controllers are compared in simulation. 

The FSM system operates as a dual input -- dual output (DIDO) system, where tip and tilt angles $\theta$ and $\phi$ nonlinearly map to $x$ and $y$ positions on a plane. Although the nonlinearity is minimal, it arises from the coupling of both FSM axes, wherein adjustments in one axis influence the behavior of the other. Here, the axis-coupling effect is taken to be negligible and single-axis controllers are designed assuming linearity. Hence, the FSM is broken down into 2 single input -- single output (SISO) systems, which was shown to be a good approximation in simulation.\\

\subsection{P Control}
The SuperBIT FGS uses a proportional term (P) that multiplies the error on the centroid state estimate \(x_k\) in each axis by a gain \(K_{fgs}\). As laid out in Romualdez \cite{Romualdez2018}, the centroid state is defined by sensor fusion of FSCs and gyros. This approach helps mitigate the latency in raw centroid measurements by generating a pseudo-measurement, which is used to correct the centroid state and covariance. From the centroid state estimates, the tip-tilt command is a simple geometric relationship given by

\begin{equation}
    \phi_{x,k} = K_{fgs} (x_k - x_{k,d})=K_{fgs}\ e_k
\end{equation}

\noindent where \(K_{fgs}\) is the proportional gain that relates centroid location to the tip-tilt angle \(\phi_{x,k}\) in the direction of \(x_k\). This method is fairly simple, and is used as a baseline to compare other methods.\\

\subsection{PID Control}
A PID controller was tuned for each FSM axis. The PID gains multiply the error on the centroid position, its derivative and integral, to send commands to the FSM. In the time domain,

\begin{equation}
    u(t) = K_p\ e(t) + K_d\ \dot{e}(t) + K_i \int_{0}^{t}e(\tau)\,\mathrm{d}\tau\ ,
\end{equation}

\noindent and in the frequency domain, 
\begin{equation}
    C(s) = K_p + K_d s + K_i \frac{1}{s} 
\end{equation}

\noindent where $K_d$, $K_i$ and $K_p$ are the derivative, integral and proportional gains to be tuned, respectively.\\

\subsection{LQR Control}

The Linear Quadratic Regulator (LQR) is formulated to restore a system to equilibrium or to maintain it at equilibrium in the presence of disturbances. The system, as modeled in a preceding section, serves as the basis for determining the optimal gains for the LQR approach. The matrix of optimal linear-quadratic gains (\(\textbf{K}\)) is computed to minimize the cost function 
\begin{equation}
    J = \int_0^\infty (\textbf{x}^\intercal \textbf{Q} \textbf{x} + \textbf{u}^\intercal \textbf{R} \textbf{u}) \ dt
\end{equation}
\noindent using the control law 
\begin{equation}
    \textbf{u} = -\textbf{K}\textbf{x} \ .
\end{equation}

\noindent $\textbf{K}$ is computed from

\begin{equation}
    \textbf{K} = -\textbf{R}^{-1}\textbf{B}^\intercal \textbf{S}
\end{equation}
\noindent where $\textbf{S}$ is the solution to the Algebraic Riccati Equation (ARE). The $\textbf{Q}$ and $\textbf{R}$ matrices, matrices that weight the relative importance of the state and input, were then adjusted to achieve the smallest possible error with an acceptable control effort.\\

\subsection{Simulation Results}

Using the model defined earlier with FGS flight data disturbance added, all three controllers are compared for an exposure time of 300 seconds. 

\begin{figure}[H]
    \centering
    \includegraphics[width=0.83\textwidth]{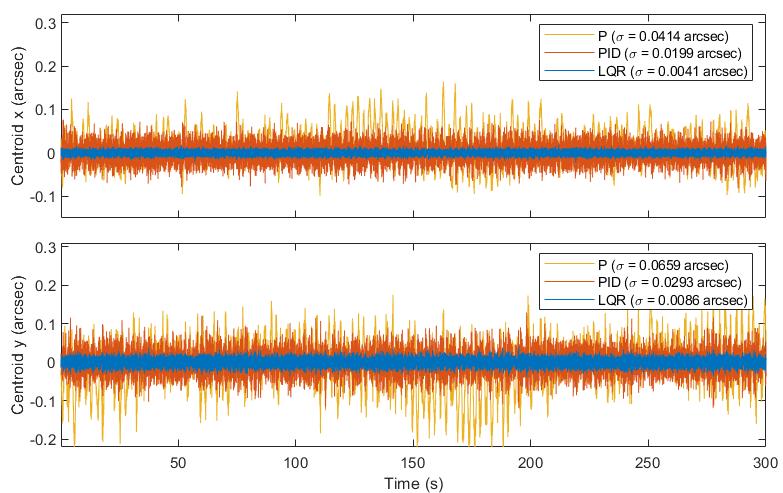}
    \caption[]{Resultant focal plane jitter: Controlled centroid position for proportional control, PID control and LQR control.}
    \label{fig:controlled_centroid}
\end{figure}

Figure \ref{fig:controlled_centroid} shows that the LQR controller is the most effective at managing jitter, followed by the PID controller, and then the P controller. The input jitter of $200$ mas in the $x$-axis and $400$ mas in the $y$-axis is reduced by the LQR controller to $4$ mas and $8$ mas, respectively, representing a reduction of nearly $98\%$ in both axes. The PID controller also achieves significant reductions, lowering jitter to $20$ mas in the $x$-axis and $29$ mas in the $y$-axis. The P controller reduces jitter as well, but to a lesser degree than the LQR and PID controllers. The input jitter being more significant in $y$ than in $x$ does not seem to affect the performance of the controllers, with the reduction being mostly proportional in both axes. 

The better performance of the LQR controller comes at a cost: a higher control effort. LQR commanded angles that are significantly larger than those of the P and PID controllers, yet still well within the $-2$ to $2$ mrad angular range of the tip-tilt actuator. Limitations to the simulation results include the use of a first-principles model rather than system identification, the need for validation against real-world scenarios and specific hardware parameters and the robustness of the control system against varied disturbances, all of which will be completed in future iterations of this integrated modeling approach. Overall, from the simulation results and considering the scaling to GigaBIT as well as the limitations of the methodology, the centroid stability requirement was set to 0.03 arcseconds RMS. This value is a compromise, chosen to be in the same order of magnitude as the standard deviations from Figure \ref{fig:controlled_centroid}, while also allowing for some margin due to the limitations of the simulation.

\section{GigaBIT FGS Requirements}

The flight data, modeling and control analysis of the previous sections directly drives into the requirements for the GigaBIT optics. This work is part of the broader GigaBIT Phase A study and Request for Proposals (RFP) issued in 2024. GigaBIT is designed to advance our understanding in key areas of astrophysics and cosmology, specifically in studies of globular clusters, stellar streams, gravitational lensing, and rapid follow-up of gravitational wave events. In continuation of its predecessor, SuperBIT, GigaBIT retains the same scientific objectives, which include studying galaxy clusters by measuring the weak lensing effects on background galaxies and probing cosmological parameters through observations of these clusters. Additionally, GigaBIT is designed to function as a general-purpose instrument, enabling scientists to investigate a diverse range of scientific cases. These science objectives drive GigaBIT's entire design and motivate the mission-level requirements. For brevity, only the driving Optical Telescope Assembly (OTA) and Fine Guidance System functional and engineering requirements are listed below, all of which directly flow down from the science objectives.

\begin{table}[H]
\caption{Driving GigaBIT OTA and FGS requirements.}
\centering
\begin{tabularx}{\textwidth}{lX} % Use tabularx with \textwidth and specify column widths
\hline 
Req \# & Requirement \\ \hline
GB-OTA-010 & Operating Conditions: Maintain optical performance throughout night at float, and over an elevation range of 20 to 60 degrees.\\
GB-OTA-020 & Science camera active field of view (FOV): 0.18 degrees squared. \\
GB-OTA-030 & Wavelength Coverage: The observatory spectral coverage shall extend from 300 nm to 800 nm. \\
GB-OTA-050 & Meet or exceed the following Point Spread Function (PSF) full width at half maximum (FWHM) benchmark values across the FOV of the science camera: \begin{center}
\begin{tabular}{|c|c|}
\hline
\textbf{Wavelength Range (nm)} & \textbf{PSF FWHM (arcseconds)} \\
\hline
300 to 350 & $\leq 0.075$  \\
\hline
350 to 400 & $\leq 0.08$ \\
\hline
400 to 600 & $\leq 0.09$\\
\hline
600 to 800 & $\leq 0.13$ \\
\hline
\end{tabular}
\end{center} \\
GB-OTA-060 & Strehl ratio greater than 0.85 over the science camera FOV. \\
GB-OTA-070 & Throughput of optical system to exceed 0.40 across the full wavelength range and FOV. \\
GB-FGS-010 & Stability: LOS pointing error of less than 0.03 arcseconds RMS in both science camera axes.\\
GB-FGS-020 & Meet pointing requirement for exposures of at least 310 seconds.\\
\hline
\end{tabularx}
\label{tab:SimulationLimitations}
\end{table}

The FGS design for GigaBIT is driven by key requirements, such as the 0.03 arcsecond LOS pointing error, which informs actuator specifications and control design. Additionally, the resolution and Strehl ratio are crucial for maintaining optical performance and achieving high image quality. Throughput is essential for maximizing the signal-to-noise ratio, especially for faint objects, and the science camera's FOV ensures wide-area imaging capabilities. These requirements are fundamental to fulfilling GigaBIT's scientific objectives and ensuring mission success, and will be driving factors for upcoming stages of the design process.

\section{CONCLUSION}
This paper provides an integrated modeling analysis to establish the requirements for the FGS, focusing on fine pointing and jitter management. The analysis utilized flight data to motivate the design of the FGS, gaining insights into the operating environment and informing the choice of sensors. Various control strategies were compared, and both Linear Quadratic Regulator (LQR) and Proportional-Integral-Derivative (PID) controllers were found to be effective for managing jitter. These steps are part of the broader GigaBIT Phase A study, which emphasizes precise pointing for imaging purposes. The modeling and analysis conducted in this study are critical for defining the functional and engineering requirements of the FGS and ensuring it meets the stringent performance criteria. A GigaBIT test flight is scheduled for the next few years to validate these design requirements and system performance, further contributing to the mission's objectives in advancing astrophysical and cosmological research.

\section{ACKNOWLEDGMENTS}

 The SuperBIT experiment was developed as a collaboration between the University of Toronto, Princeton University, Durham University, and the Jet Propulsion Laboratory. SuperBIT is supported in Canada by the Canadian Institute for Advanced Research (CIFAR), the Natural Science and Engineering Research Council (NSERC), and the Canadian Space Agency (CSA). The US team acknowledges support from the NASA APRA grant 80NSSC22K0365. British team members acknowledge support from the Royal Society (grant RGF/EA/180026), the UK Science and Technology Facilities Council (grant ST/V005766/1), and Durham University’s astronomy survey fund. GigaBIT is additionally supported by the Canadian Foundation for Innovation (CFI) and the Ontario Research Fund (ORF).

% References
\bibliography{report} % bibliography data in report.bib
\bibliographystyle{spiebib} % makes bibtex use spiebib.bst

\end{document}